\documentclass{aa}
\usepackage{graphicx,caption,color}
\usepackage{txfonts}
\usepackage{hyperref}
\usepackage{amsmath}
\usepackage{textcomp}

\usepackage{gensymb}
\usepackage{subcaption}
\begin{document}

\title{A search for transiting planets in the $\beta$ Pictoris system
  \thanks{Based on data collected by the BRITE-Constellation
    satellite mission built, launched, and operated thanks to support from the Austrian Aeronautics and Space Agency and the University of Vienna, the Canadian Space Agency (CSA), and the Foundation for Polish Science \& Technology (FNiTP MNiSW) and National Centre for Science (NCN).}}

   \author{M. Mol Lous\inst{1}
   \and E. Weenk\inst{1}
   \and M.A. Kenworthy\inst{1}
        \and K. Zwintz\inst{2}\fnmsep\thanks{Elise Richter Fellow of the Austrian Science Funds (FWF).}
        \and R. Kuschnig\inst{3}
   }

   \institute{ {Leiden Observatory, Leiden University, P.O. Box 9513, 2300 RA Leiden, The Netherlands}
   \and {Institut f{\"u}r Astro- und Teilchenphysik, Universit{\"a}t Innsbruck, Technikerstrasse 25, 6020 Innsbruck, Austria} 
   \and{Graz University of Technology, Institute of Communication Networks and Satellite Communications, Infeldgasse 12, 8010 Graz, Austria} }
   
 \date{Received 13 September 2017 / Accepted 22 March 2018}

  \abstract
   {Transiting exoplanets provide an opportunity for the
characterization of their atmospheres, and finding the brightest star in the sky with a transiting planet
enables high signal-to-noise ratio observations.
The Kepler satellite has detected over 365 multiple transiting exoplanet
systems, a large fraction of which have nearly coplanar orbits.
If one planet is seen to transit the star, then it is likely that other planets in the system will transit the star too.
The bright $(V=3.86)$ star $\beta$ Pictoris is a nearby young star with a debris
disk and gas giant exoplanet, $\beta$ Pictoris b, in a multi-decade orbit
around it.
Both the planet's orbit and disk are almost edge-on to our line of sight.
   }
   {We carry out a search for any transiting planets in the $\beta$ Pictoris system
with orbits of less than 30 days that are coplanar with the planet $\beta$ Pictoris b.
  }
   {We search for a planetary transit using data from the
\emph{BRITE-Constellation} nanosatellite \emph{BRITE-Heweliusz},
analyzing the photometry using the Box-Fitting Least Squares Algorithm
(BLS).
The sensitivity of the method is verified by injection of
artificial planetary transit signals using the Bad-Ass Transit Model
cAlculatioN (BATMAN) code.}
   {No planet was found in the \emph{BRITE-Constellation} data set.  We
rule out planets larger than 0.6 $\mathrm{R_J}$ for periods of less than 5
days, larger than 0.75 $\mathrm{R_J}$ for periods of less than 10 days, and
larger than 1.05 $\mathrm{R_J}$ for periods of less than 20 days.}
   {}


   \maketitle
 

\section{Introduction}
\label{sec:introduction}

The discovery of the first extrasolar planets over twenty years ago
through the radial velocity technique \citep{Mayor95,Marcy95} rapidly
led to the detection of multiple planet systems
\citep{Butler99,Fischer02,Marcy01}.
As the number of exoplanet detections reached into the hundreds, the
architecture of these systems began to show a wide range of diversity in both
their semimajor  axis distribution and their eccentricities.
The unexpected detection of giant planets with semimajor axes smaller
than the orbit of Mercury in our own solar system rapidly led to the
first detection of a transiting planet \citep{Charbonneau00,Brown01}
and the subsequent characterization of its atmosphere
\citep{Charbonneau02}.
The {\it Kepler} satellite led to the discovery of over 2000
exoplanets, along with over 365 multiple-planet systems where the
mutual inclinations of the planets is in the range of 1 to 2 degrees
\citep{Fabrycky14}.

The signal-to-noise ratio of a transiting planet is partly driven by the
apparent brightness of the host star, and so several searches are
underway to find the brightest star in the sky with a transiting
planet.
Several transit searches therefore focus on the brightest stars,
including
the Multi-site All-Sky CAmeRA (MASCARA), which surveys the sky from the ground
for stars with magnitudes from 4 to 8 \citep{Talens17}, and 
the Kilodegree Extremely Little Telescope (KELT), which searches for transits from
the ground for magnitudes from 8 to 10 \citep{Pepper07}.
There are also future projects for measuring transits around bright
stars via space telescopes.
The Transiting Exoplanet Survey Satellite (TESS) is planned for launch in
mid-2018 and will look at stars 10--100 times as bright as {\it Kepler}
\citep{Ricker14}.
Further in the future is the  PLAnetary Transits and Oscillations of
stars (PLATO).
Planned for launch in 2024 it will, among other things, look for terrestrial
planets around stars similar to our
Sun\footnote{http://sci.esa.int/plato/}.
Looking for the brightest star with a transiting Earth-like planet is
of great importance, as this would make it an ideal target for further
investigation with a telescope such as the James Webb Space Telescope.
The present-day interest in Earth-like exoplanets is confirmed by the
attention that the TRAPPIST-1 planetary system received following its
discovery in 2017 \citep{Gillon17}.

The $\beta$ Pictoris system is a young \citep[$\sim$23
  Myr;][]{Mamajek14} and nearby \citep[19.3 pc;][]{Leeuwen07} system, which consists of the star, a debris disk with several
kinematical components, and one gas giant planet ($\beta$ Pictoris b).
The star is known to be a $\delta$ Scuti pulsator: three pulsation
frequencies were discovered by \citet{Koen03a} using ground-based
photometric time series.
Shortly after that, \citet{Koen03b} reported the presence of 14
pulsation frequencies found from dedicated spectroscopic time series.

In 1983, IRAS detected excess infrared material around $\beta$
Pictoris \citep{Aumann85}.
A year later, a circumstellar disk was observed optically around the
star by \citet{Smith84}.
The disk is nearly edge-on and its outer extent varies from 1450 to
1835 AU \citep{Larwood01}.
Distinct belts around $\beta$ Pictoris have been detected at distances
of 14, 28, 52, and 82 AU \citep{Wahhaj03}, kinematically suggesting the
presence of planets in the system.
The inner disk of $\beta$ Pictoris is warped, implying the presence of
a secondary companion inside the disk \citep{Mouillet08}.
The stellar and planetary parameters for $\beta$ Pictoris and its
planet are listed in Table~\ref{table:betapic}.

\begin{figure}
\centering
\includegraphics[width=0.5\textwidth]{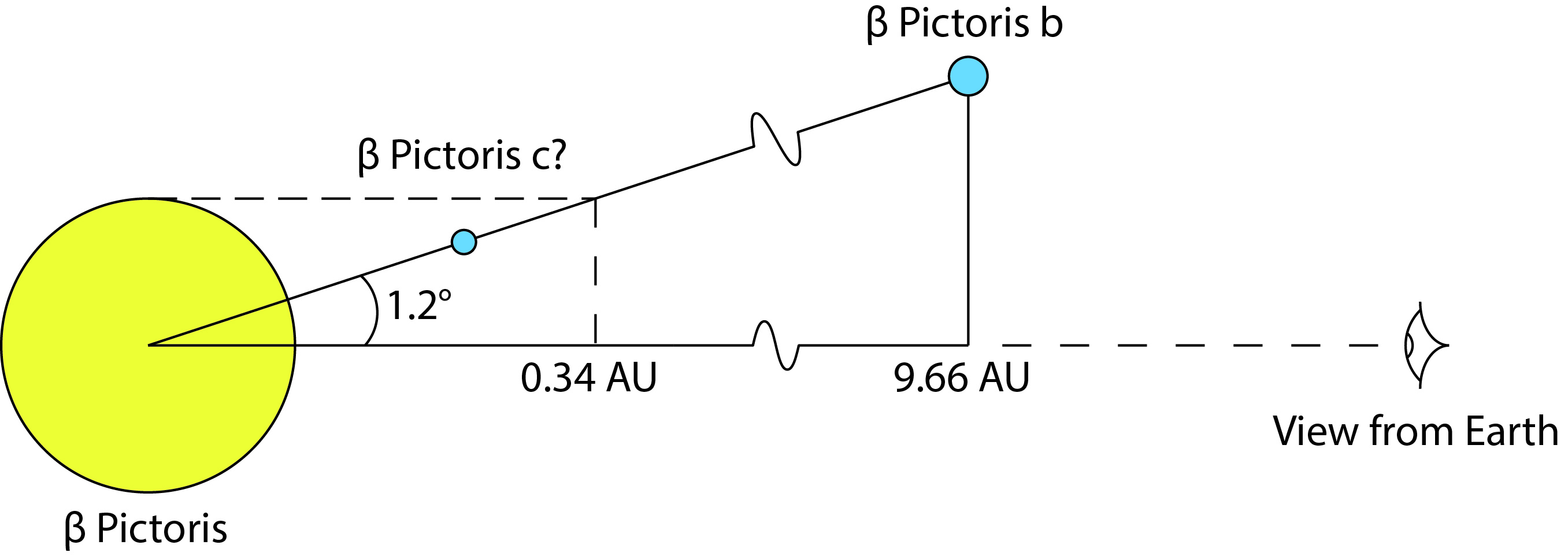}
\caption{Image of the $\beta$ Pictoris system showing the geometry of
$\beta$ Pictoris b and a coplanar third planet interior to b.
A planet coplanar with $\beta$ Pictoris b $i=88.81^{\circ}$ will transit
along diminishing chords of the star out to a radius of 0.34 au,
corresponding to an orbital period of 54 d.}
\label{fig:betapic}
\end{figure}

A planet was directly imaged in 2009 \citep{Lagrange09,Lagrange10} at
a wavelength of 3.5 $\mu m$ by the NACO camera on the VLT at Paranal.
The planet $\beta$ Pictoris b will not transit its host star  because
of its distance from the star and its inclination of $88.81^{\circ}
\pm{0.12}^{\circ}$ as seen from Earth \citep{Wang16};  see
Figure~\ref{fig:betapic}.
Planets form from a natal circumstellar disk, and therefore are
coplanar with coaligned orbital angular momentum vectors, as seen in
multiple transiting systems in {\it Kepler} and the solar system
\citep{Fabrycky14}.
The high inclination of $\beta$ Pictoris and the morphology of the
disk suggests that if there are any other planets in the $\beta$
Pictoris system then they will also have high inclinations and
possibly transit the stellar disk.

We examine photometric data from the BRITE-Constellation
nanosatellite mission, which observed $\beta$ Pictoris for $\delta$
Scuti pulsations in 2015.
In Section~\ref{sec:data} we describe the BRITE data set, and
Section~\ref{sec:BLS} is dedicated to the analysis we use on this
data.
The results and conclusions are discussed in
Section~\ref{sec:discussions} and \ref{sec:conclusions}.

\section{Data}
\label{sec:data}
   \subsection{BRITE-Constellation}
   \label{subsec:BRITE}

The data for this study was obtained using BRITE-Heweliusz (BHr), one
of the \emph{BRIght Target Explorer (BRITE)}-Constellation
nanosatellites.
\emph{BRITE-Constellation} \citep{Weiss14,Pablo16,Popowicz16} is a
fleet of nanosatellites (6 launched; 5 operational) operated by a consortium
consisting of Austria, Canada, and Poland).
BHr carries a red filter \citep{Weiss14}, and has a nearly circular
Sun-synchronous orbit with an altitude of approximately 635 km and a
period of 97.1 minutes.

\subsection{\texorpdfstring{$\beta$}{Beta} Pictoris data} \label{subsec:pictorisdata} 

\begin{figure*}
\centering
\includegraphics[width=1.0\textwidth]{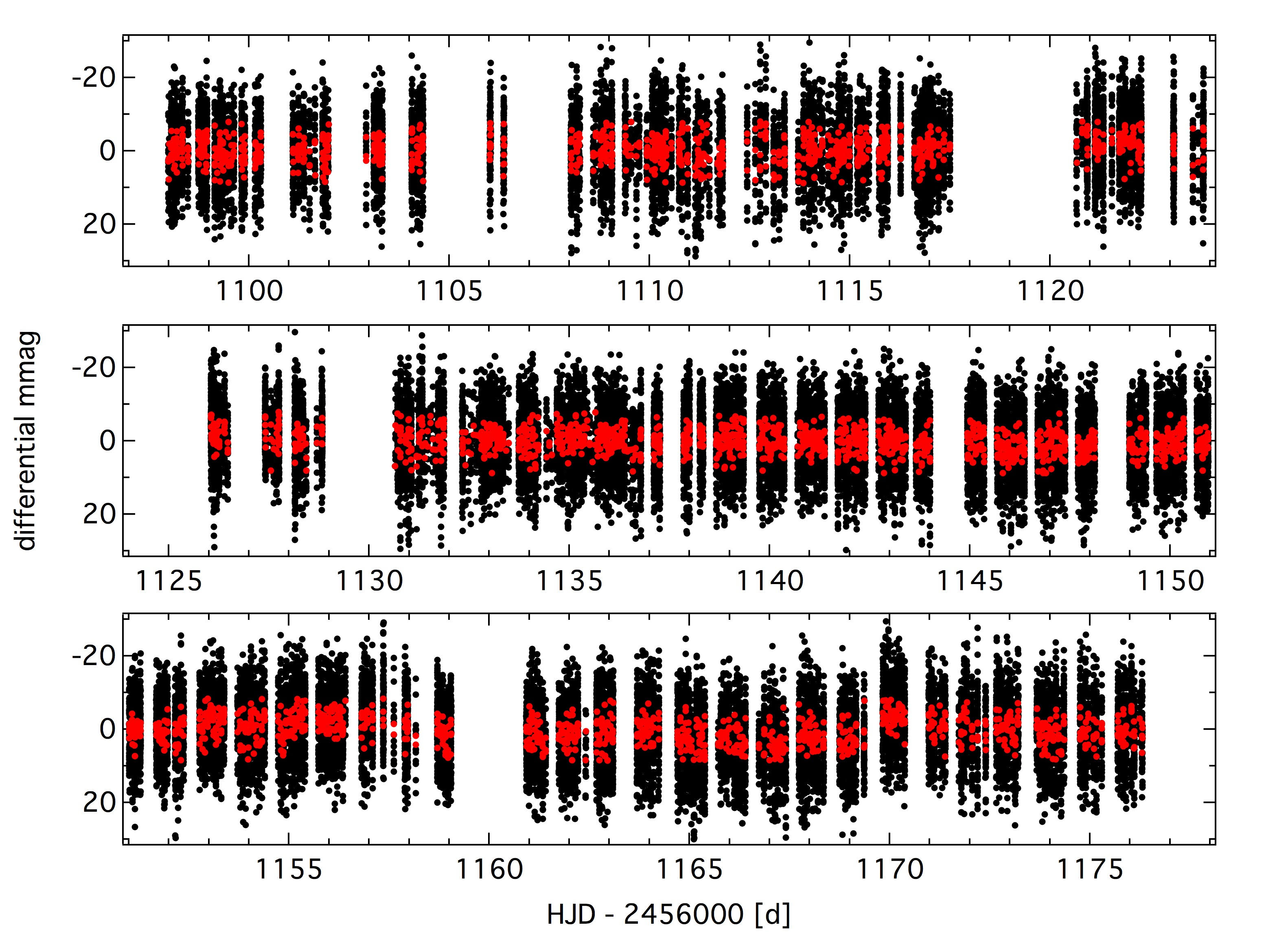}
\caption{Light curve of the detrended photometry from the BRITE
satellite for $\beta$ Pictoris over the 78-day run. The original detrended data are shown in black, data binned over three-minute intervals are shown in red. Each subset covers a time span of 27 days.}
\label{fig:brite_light}
\end{figure*}

Observations were made between 2015 March 16 UT and 2015 June 02 UT
for a total of 78 days.
The original data set, which can be retrieved from the BRITE Public Data
Archive,\footnote{\url{https://brite.camk.edu.pl/pub/index.html}} 
consists of 47330 individual photometric measurements obtained in
stare mode \citep{popowicz2017}.
After main data reduction pipeline processing \citep{popowicz2017},
the light curve was decorrelated using the steps described in
\citet{pigulski2016} and points with excessive photometric deviations
and noise were removed.
The final data set contains 44246 measurements (see
Figure~\ref{fig:brite_light}).

The frequency analysis of the BRITE photometric time series was
performed using the software package Period04 \citep{lenz2005}, which
combines Fourier and least-squares algorithms.
Frequencies were then pre-whitened and considered to be significant it
their amplitudes exceeded $4\times$ the local noise level in the
amplitude spectrum \citep{breger1993,kuschnig1997}.
We identified eight $\delta$ Scuti pulsation frequencies in the range
between 32 and 61 d$^{\rm -1}$ with amplitudes between 0.3 and 1.4
mmag, which were then subtracted from the data.
The residual light curve was normalized to unity and yields a standard
deviation of 0.86\%.
A detailed discussion of the pulsational content of $\beta$ Pictoris
is summarized in \citet{Zwintz17} and will be described fully together
with a detailed asteroseismic interpretation in Zwintz et al. (in prep).

   \begin{table}
      \caption[]{Stellar and planetary parameters for $\beta$ Pictoris}
         \label{table:betapic}
      \begin{tabular*}{\linewidth}{@{\extracolsep{\fill}}lll}
      \hline
      \hline
      \noalign{\smallskip}
                Parameter & Value & Source \\
          \hline
          \noalign{\smallskip}
                Total mass ($M_T$)      & $1.80^{+0.03}_{-0.04}\ \mathrm{M_{\odot}}$  & \citet{Wang16}\\ \noalign{\smallskip}
                Radius $\beta$ Pic ($R_*$) & $1.53\ \mathrm{R_{\odot}}$   & \citet{Wang16}\\       \noalign{\smallskip}
        Distance ($d$) & 19.3 pc & \citet{Leeuwen07}\\      \noalign{\smallskip}
                Radius $\beta$ Pic b ($R_p$) & $1.46\pm 0.01\ \mathrm{R_J}$ & \citet{Chilcote17}\\      \noalign{\smallskip}
                Inclination $\beta$ Pic b ($i$) &$88.81^{\circ} \pm{0.12}^{\circ} $  &\citet{Wang16} \\  \noalign{\smallskip}
        Semimajor axis ($a$) & $9.66^{+1.12}_{-0.64}$ AU &\citet{Wang16}\\ \noalign{\smallskip}
        \hline
                \end{tabular*}
   \end{table}

\section{Analysis and results}
\label{sec:BLS}

We use the Box-Fitting Least Squares (BLS) algorithm to perform a
search for transiting planets in the BRITE data.
BLS is an algorithm \citep[designed by][]{Kovacs02} that analyzes a
photometric time series and searches for a flat bottomed planetary
transit, returning the most probable orbital periods and transit
depth.
For this study, the ``eebls''
routine\footnote{\url{https://github.com/dfm/python-bls}} was 
used.
These are \verb+Python+ bindings for the original \verb+Fortran+
implementation of BLS from \citet{Kovacs02}, modified for edge effects
when the transit event happens to be divided between the first and
last bins, as the original BLS yielded lower detection efficiency in
these cases.

We first determined the parameters using an artificial data set with
the same cadence and noise properties as the BRITE data set.
Transits were injected into this artificial data and the rate of
recovery was determined from many random trials.

\subsection{Determining the BLS parameters}
\label{subsec:BLSparameters}
    
A number of parameters can be set for the eebls routine: 1) $nf$, the
number of frequency points in which the spectrum is computed; 2)
$f_{min}/f_{max}$, the minimum/maximum frequency; 3) $df$, the
frequency step size; 4) $nb$, the number of bins in the folded time
series at any test period; 5) $qmi/qma$, the minimum/maximum
fractional transit length to be tested.
The values of these parameters used in this study can be seen in
Table~\ref{table:bls}.

     \begin{table}[h]
      \caption[]{Parameter values in the eebls routine}
         \label{table:bls}
      \begin{tabular*}{\linewidth}{@{\extracolsep{\fill}}lllllll}
      \hline
      \hline
      \noalign{\smallskip}
      $nf$ & $f_{min}$ & $f_{max}$ & $df$ & $nb$ & $qmi$ & $qma$ \\
      $1000$ & $0.033333$ & $1$ & $0.00097$ & $1750$ & $0.005$ & $0.12$ \\
      $16000$ & $0.033333$ & $1$& $0.00006$ & $1750$ & $0.005$ & $0.12$ \\
      \noalign{\smallskip}
      \hline
      \hline
      \end{tabular*}
   \end{table}

A value of 1000 and 16000 was used for $nf$ for the majority of the
simulations.
The minimum frequency $f_{min}$ was chosen as $1/P_{max}$, the maximum
period to test for, and in the same way $f_{max} = 1/P_{min}$.
The value of $df$ was calculated from the $nf$ and $f_{min}$ (and
$f_{max}$) values using the following formula: $$ df = \frac{f_{max}
  -f_{min}}{nf}$$ The value for $nb$ is 1750, which allows for 25 data
points per bin for the longest trial period.
For $qmi$ and $qma$ we used the following equation to obtain transit
durations:

\begin{equation}
\label{eq:transitduration} T_{time} =
\frac{P}{\pi} \mathrm{sin}^{-1}\left(\frac{\sqrt{(R_* + R_p)^2 -
(bR_*)^2}}{a}\right)
\end{equation}
with

\begin{equation}
\label{eq:semiaxis} a^3 = P^2 \frac{G(M_*+m_p)}{4\pi^2}, \end{equation}where $P$ is the period of the planet and $b$ the impact parameter.
A planet orbiting $\beta$ Pictoris will have a transit duration of 7.5
hours for a 30-day period, corresponding to a transit fraction of
about 0.01, dropping to a transit period of 2.4 hours for a one-day
orbital period, corresponding to a transit fraction of 0.1.
To make sure the values of $qmi$ and $qma$ covered the period range of
this search, they were set to $qmi=0.005$ and $qma=0.12$.

\subsection{Sensitivity plots}
\label{subsec:sensitivity}

  To test the sensitivity of the BLS algorithm in recovering transits
  from the \emph{BRITE} data, fake transits were inserted into the
  data set and then recovered.
These transits were simulated with the \verb+Python+ package
\verb+BATMAN+ \citep[Bad-Ass Transit Model
  cAlculatioN;][]{Kreidberg15}.
With \verb+BATMAN+ we simulate the transit curve given a planet radius
$R_p(i)$ and periods $P(i)$, along with the stellar mass and radius
and the orbital inclination (listed in Table~\ref{table:betapic}).
A nonlinear limb darkening model was used.

\subsubsection{White noise}

The time data points and standard deviation of the flux from the
\emph{BRITE} data were used to generate a  data set similar to the
\emph{BRITE} data with Gaussian noise.
The \verb+BATMAN+ curves were inserted into this generated data set 
and the BLS algorithm applied to generate a periodogram.
The highest peak in the periodogram was given a rank $k=0$, the second
highest peak $k=1$, the third highest peak $k=2$, and so on.
If the period $P_{k=0}$ was within 5\% of the inserted period $P(i)$
(so that $\left|P_{k=0} - P(i) \right| < 0.05 P(i)$), it was
considered a successful retrieval.
For each trial of $P$ and $R$, a new transit model was injected into
the data with the transit time $T$ randomized over the interval $0$ to
$P$.
The rank of the period peak that corresponds to the injected orbital
period $P$ was noted.
The mean rank $\bar{k}$ was calculated as

$$ \bar{k} = \sum_{all n} k_n / n, $$
where $\bar{k}=0$ means that the correct period was identified in
all the trials.

The success rate of this retrieval told us whether or not we would be
able to find planets with radius $R_p(i)$ and period $P(i)$ in the
\emph{BRITE} data.
The period was varied between 1 and 30 days in steps of 0.25 days.
The 30-day upper limit was set to ensure three transits detectable
within the 78-day window.

\begin{figure}
\centering
\includegraphics[width=0.6\textwidth]{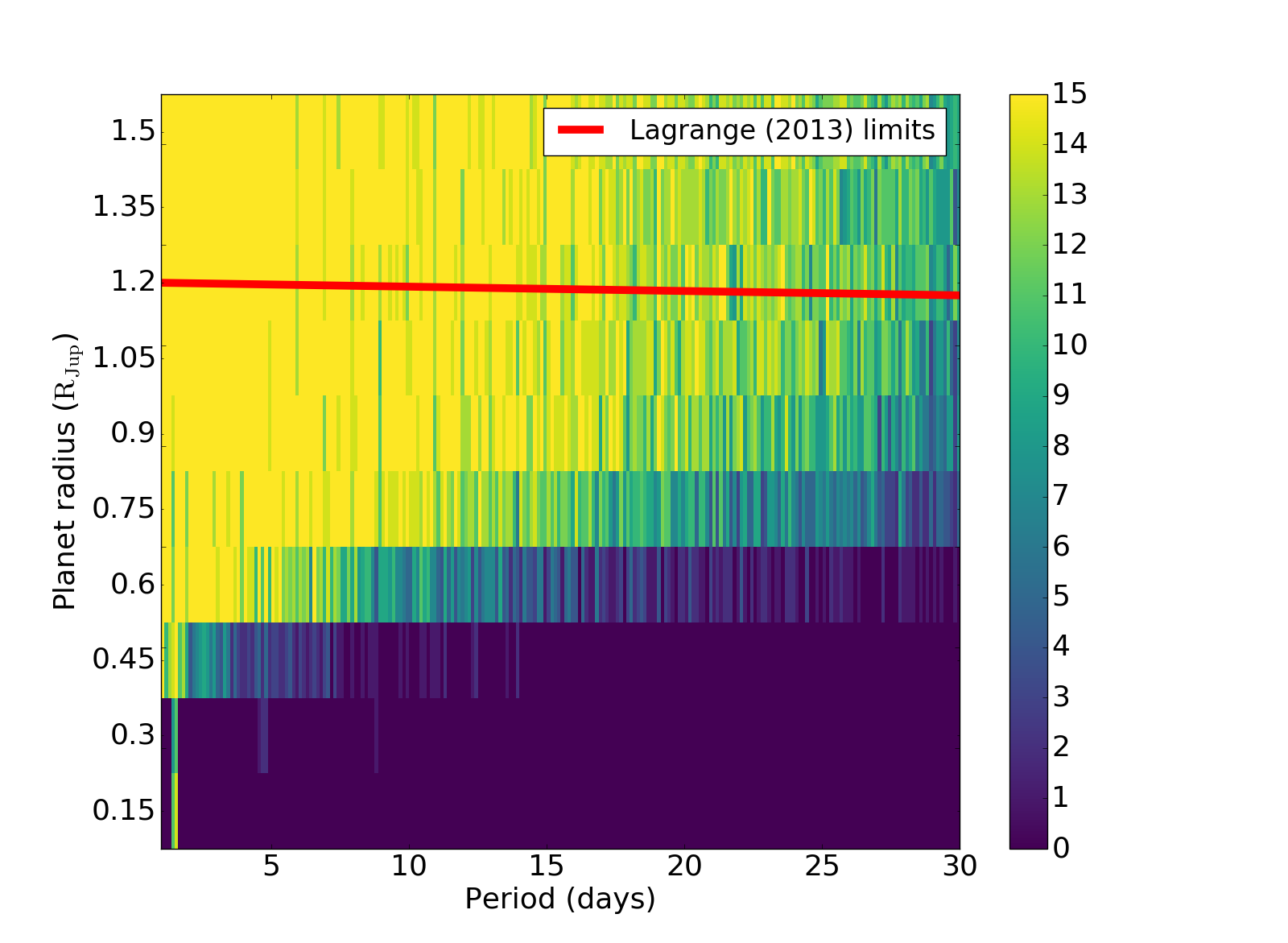}

\caption{Sensitivity map for $\beta$ Pictoris using simulated Gaussian
  noise. The red line indicates the detection limit  set by
  \citet{Lagrange13}.}

\label{fig:sensitivitymap49}
\end{figure}

The model radius was varied between 0.15 and 1.5 $R_{J}$ in steps of
0.15 $R_{J}$.
The upper limit for this radius was based on the \citet{Lagrange13}
radial velocity-based detection limits, where they found that for a
period of 3 days a planet of 1.5 $M_{J}$ can remain
undetected with the current available data, while for a period of 10
days this is 2.1 $M_{J}$.
The results are displayed in a sensitivity map of period $P$ and
radius $R$ (see Figure~\ref{fig:sensitivitymap49}).

\subsubsection{\texorpdfstring{$\delta$}{Delta} Scuti pulsation analysis}
\label{subsec:applyingthedata}

\begin{figure}
\centering
\includegraphics[width=1.0\columnwidth]{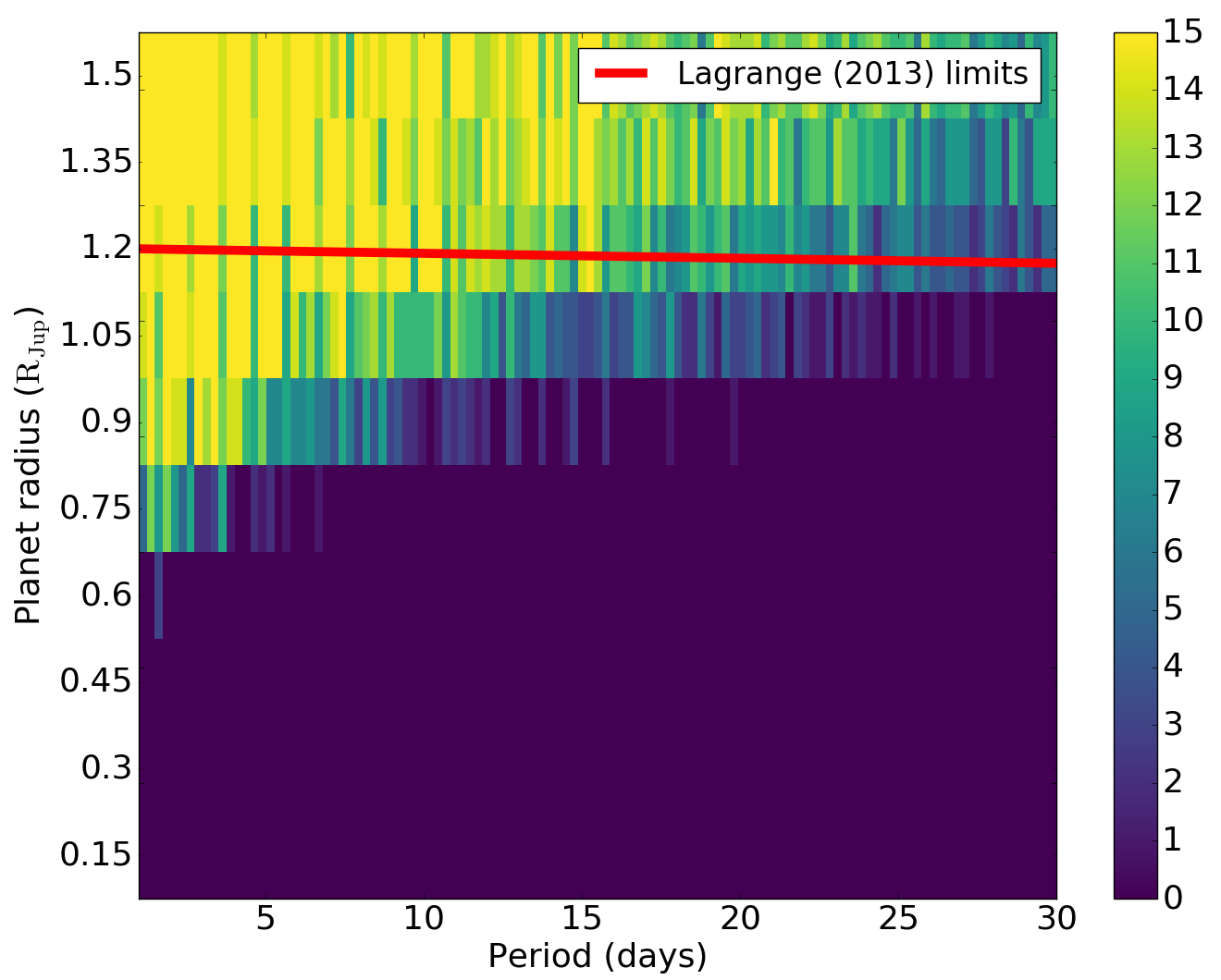}

\caption{Sensitivity map for the $\beta$ Pictoris photometry. The red
line indicates the detection limit from \citet{Lagrange13}.}

\label{fig:sensitivitymap55}
\end{figure}

The orbital period of the satellite (97.1 minutes) imposes a periodic
window function on the photometry, and this appears as spectral power
at 0.067469 days.
The $\delta$ Scuti pulsations from the star alias with this window
function and act as a source of systematic error in the photometry.
This is seen in the reduced sensitivity of the BLS analysis carried
out on the $\beta$ Pictoris photometric data seen in
Figure~\ref{fig:sensitivitymap55}.

\begin{figure}
\includegraphics[width=1.0\columnwidth]{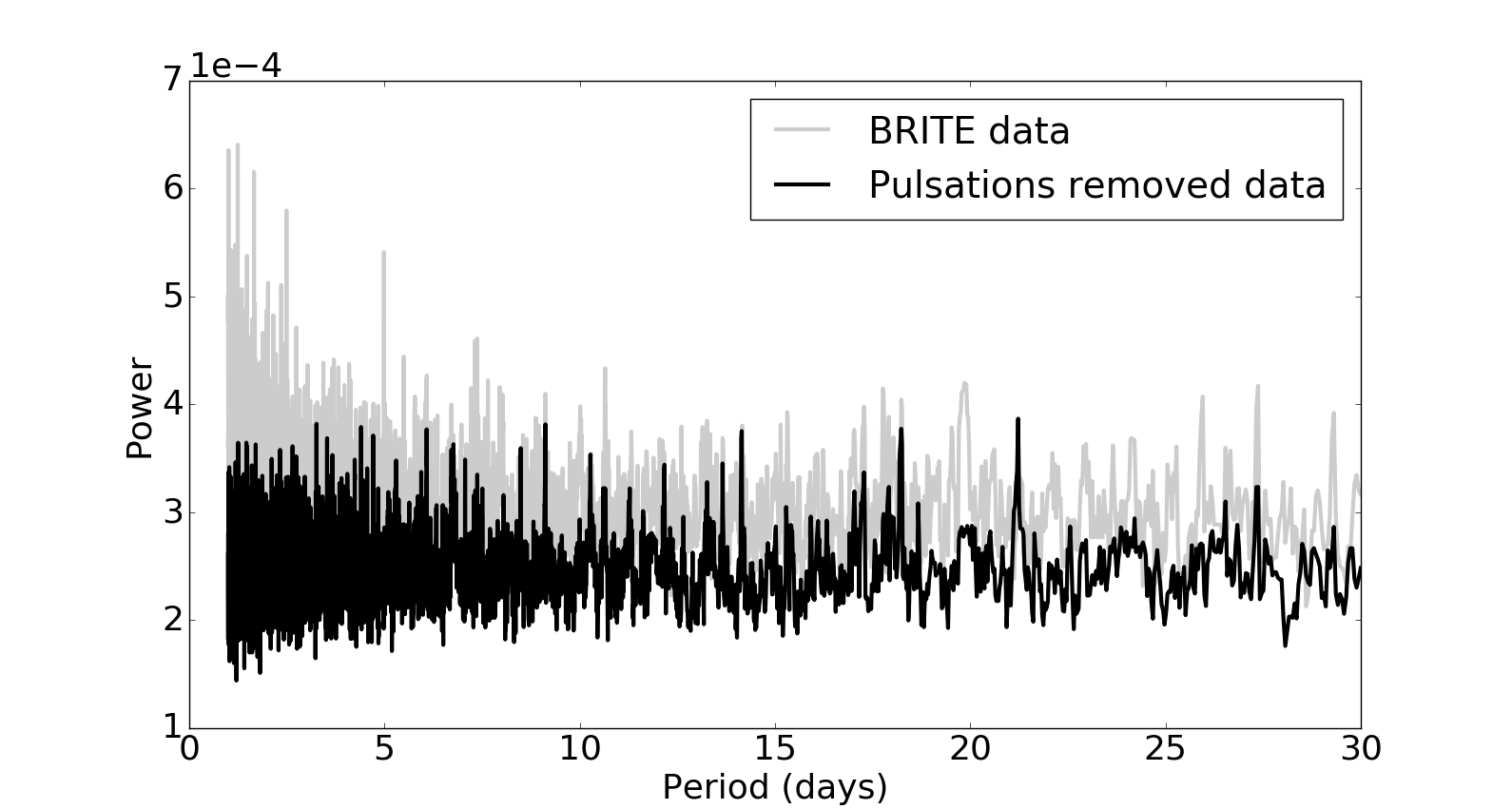}

\caption{Periodogram of the original and the pulsation removed BRITE
  data.}

\label{fig:periodograms}
\end{figure}

\begin{figure}
  \centering
  \includegraphics[width=1.0\columnwidth]{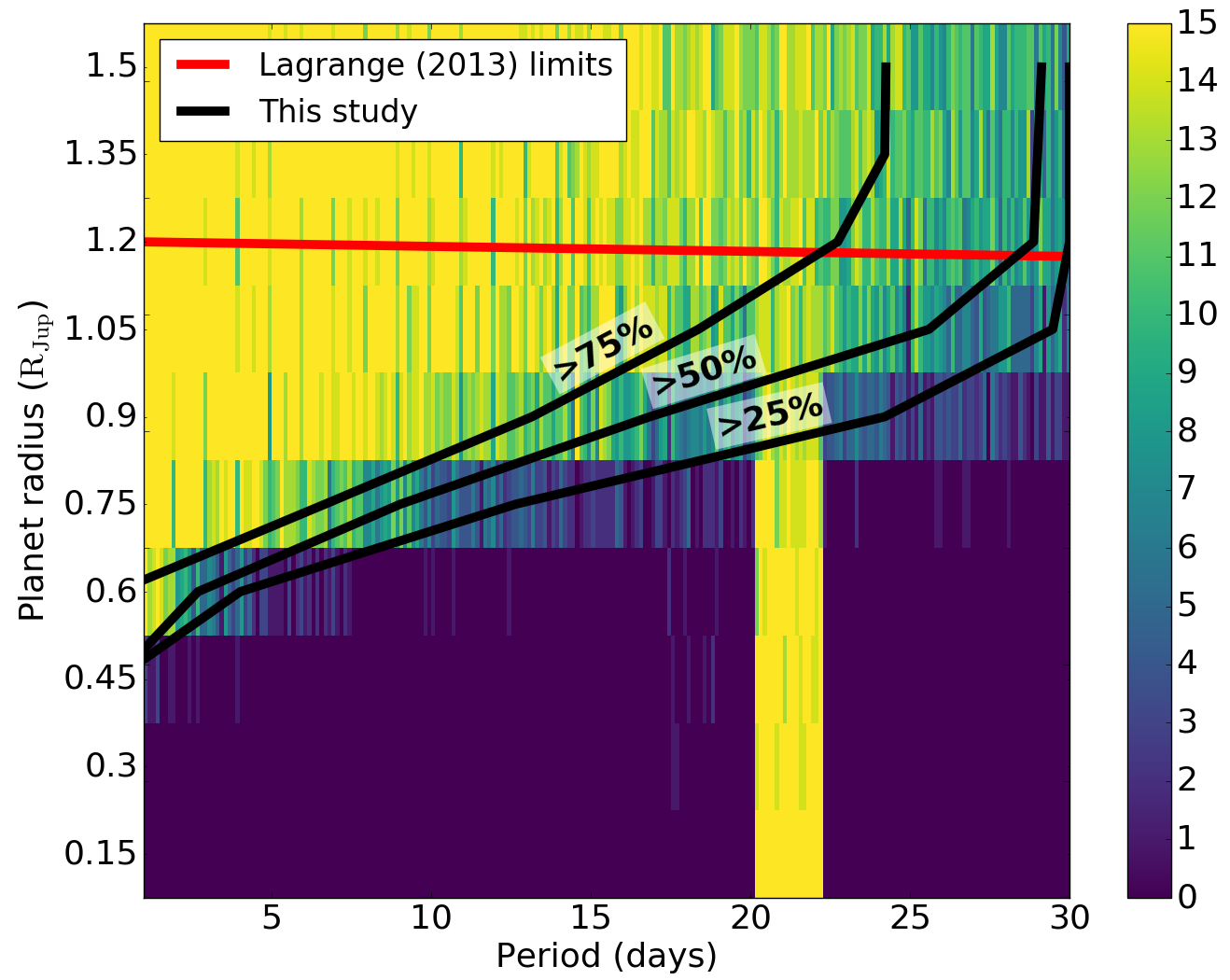}

  \caption{Sensitivity map for $\beta$ Pictoris using the pulsation
    removed data set. The red line indicates the detection limit     set by \citet{Lagrange13}, whereas the solid black lines are an
    indication of the limits set by this study.}

  \label{fig:sensitivitymap63}
\end{figure}

The $\delta$ Scuti pulsations are removed according to the
prescription in \citet{Zwintz17}.
The subtraction of the pulsations results in a decrease in power
across the periodogram, as seen in Figure~\ref{fig:periodograms}.
The amplitudes of several peaks are seen to decrease and several drop
below the mean noise floor of the data.
The transit injection analysis is repeated for this data set.
A visual inspection of the period folded light curves for the ten
highest ranked periods was made.
No plausible transiting candidate is present in the data.
To determine the upper limits of the sensitivity of this analysis, a
sensitivity map of this data is made to determine which combinations
of period and radius can be excluded.
The result is shown in Figure~\ref{fig:sensitivitymap63}.

\section{Discussion}
\label{sec:discussions}

A radial velocity study by \citet{Lagrange13} constrained possible
massive companions to $\beta$ Pictoris, placing upper limits on the
mass of any companions.
To evaluate the currently existing limits, we converted the mass limits
into limits for the radii.
We used the \verb+Python+ package called
\verb+Forecaster+\footnote{\url{https://github.com/chenjj2/forecaster}}
by \citet{Chen17}.
For a given mass, younger planets have a larger radii at younger ages
as they have a greater effective temperature.
This package is calibrated for much older planets, and it defines a
conservative lower limit on the mass.
This package takes as input the mean and standard deviation of the
mass and returns a mean and standard deviation of the radius.
Using this package, the limits for the mass given in
\citet{Lagrange13} were converted into limits for the radius.
The values we calculated are shown in Table~\ref{table:massradius}.
The limits from \citet{Lagrange13} are shown as a red line in all the
figures.
Comparing these values with the sensitivity map of the original \emph{BRITE}
data, seen in Figure~\ref{fig:sensitivitymap55} we show a significant
increase in sensitivity for smaller orbital periods.

\begin{table}[h]
\caption[]{Mass radius relationship}
\label{table:massradius}
\begin{tabular*}{\linewidth}{@{\extracolsep{\fill}}llllll}
\hline
\hline
\noalign{\smallskip}
Period (days) & 3 & 10 & 100 & 500 & 1000 \\
 Mass limit ($\mathrm{M_J})$ & 1.5 & 2.1 & 4 & 12.4 & 16 \\
 Radius ($\mathrm{R_J})$ & 1.20 & 1.18 & 1.15 & 1.09 & 1.08 \\
\noalign{\smallskip}
\hline
\hline
\end{tabular*}
\end{table}

\section{Conclusions}
\label{sec:conclusions}

The sensitivity map of the pulsation removed BRITE data lowers the
limits of possible exoplanets as set by \citet{Lagrange13}.
As was discussed in Table~\ref{table:massradius}, the limits set by
that study were 1.2 $\mathrm{R_J}$ for 3 days, 1.18 $\mathrm{R_J}$ for
10 days, and 1.15 $\mathrm{R_J}$ for 100 days.
We find limits of 0.6 $\mathrm{R_J}$ for 5 days, 0.75 $\mathrm{R_J}$
for 10 days, and 1.0 $\mathrm{R_J}$ for 20 days.
However, because of the degeneracies in converting radii to masses,
these limits are best kept as constraints on the radius of a possible
$\beta$ Pictoris c.

With $\beta$ Pictoris b moving through inferior conjunction in 2017 \citep{Wang16},
several projects are conducting photometric and spectroscopic monitoring campaigns
looking for the signatures of circumplanetary material in the Hill sphere.
In order to provide a good baseline for this transit,
BRITE-Constellation has revisited $\beta$ Pictoris from November 4,
2016, to June 22, 2017.
\footnote{\url{http://brite.craq-astro.ca/doku.php?id=betapic-ii}}
The corresponding data are currently being reduced.
A photometric monitoring project called bRing \citep{Stuik17} has two cameras surveying $\beta$
Pictoris in South Africa and Australia.
The camera in South Africa  already started observing
in February 2017, and the one based in Australia started observing in November 2017
\footnote{\url{https://www.hou.usra.edu/meetings/abscicon2017/eposter/3321.pdf}}
\footnote{\url{http://home.strw.leidenuniv.nl/~kenworthy/beta_pic_b_hill_sphere}}.

Although such observations are intended for detecting the $\beta$ Pictoris
b Hill sphere, they can also be used as additional data for the
method used in this study.
By doubling or tripling the number of data points, the sensitivity of
the plot to smaller radii planets will increase and enable tighter
constraints on, or the detection of, smaller planets.

\begin{acknowledgements}

Based on data collected by the BRITE Constellation satellite mission,
designed, built, launched, operated, and supported by the Austrian
Research Promotion Agency (FFG), the University of Vienna, the
Technical University of Graz, the Canadian Space Agency (CSA), the
University of Toronto Institute for Aerospace Studies (UTIAS), the
Foundation for Polish Science \& Technology (FNiTP MNiSW), and the
National Science Centre (NCN).
This research made use of Astropy, a community-developed core Python
package for Astronomy \citep{2013A&A...558A..33A}.
KZ acknowledges support by the Austrian Fonds zur F\"orderung der
wissenschaftlichen Forschung (FWF, project V431-NBL).

\end{acknowledgements}

\bibliographystyle{aa}
\bibliography{mybib}

\begin{thebibliography}{41}
\expandafter\ifx\csname natexlab\endcsname\relax\def\natexlab#1{#1}\fi

\bibitem[{{Astropy Collaboration} {et~al.}(2013){Astropy Collaboration},
  {Robitaille}, {Tollerud}, {Greenfield}, {Droettboom}, {Bray}, {Aldcroft},
  {Davis}, {Ginsburg}, {Price-Whelan}, {Kerzendorf}, {Conley}, {Crighton},
  {Barbary}, {Muna}, {Ferguson}, {Grollier}, {Parikh}, {Nair}, {Unther},
  {Deil}, {Woillez}, {Conseil}, {Kramer}, {Turner}, {Singer}, {Fox}, {Weaver},
  {Zabalza}, {Edwards}, {Azalee Bostroem}, {Burke}, {Casey}, {Crawford},
  {Dencheva}, {Ely}, {Jenness}, {Labrie}, {Lim}, {Pierfederici}, {Pontzen},
  {Ptak}, {Refsdal}, {Servillat}, \& {Streicher}}]{2013A&A...558A..33A}
{Astropy Collaboration}, {Robitaille}, T.~P., {Tollerud}, E.~J., {et~al.} 2013,
  \aap, 558, A33

\bibitem[{{Aumann}(1985)}]{Aumann85}
{Aumann}, H.~H. 1985, \pasp, 97, 885

\bibitem[{{Breger} {et~al.}(1993){Breger}, {Stich}, {Garrido}, {Martin},
  {Jiang}, {Li}, {Hube}, {Ostermann}, {Paparo}, \& {Scheck}}]{breger1993}
{Breger}, M., {Stich}, J., {Garrido}, R., {et~al.} 1993, \aap, 271, 482

\bibitem[{{Brown} {et~al.}(2001){Brown}, {Charbonneau}, {Gilliland}, {Noyes},
  \& {Burrows}}]{Brown01}
{Brown}, T.~M., {Charbonneau}, D., {Gilliland}, R.~L., {Noyes}, R.~W., \&
  {Burrows}, A. 2001, \apj, 552, 699

\bibitem[{{Butler} {et~al.}(1999){Butler}, {Marcy}, {Fischer}, {Brown},
  {Contos}, {Korzennik}, {Nisenson}, \& {Noyes}}]{Butler99}
{Butler}, R.~P., {Marcy}, G.~W., {Fischer}, D.~A., {et~al.} 1999, \apj, 526,
  916

\bibitem[{{Charbonneau} {et~al.}(2000){Charbonneau}, {Brown}, {Latham}, \&
  {Mayor}}]{Charbonneau00}
{Charbonneau}, D., {Brown}, T.~M., {Latham}, D.~W., \& {Mayor}, M. 2000, \apjl,
  529, L45

\bibitem[{{Charbonneau} {et~al.}(2002){Charbonneau}, {Brown}, {Noyes}, \&
  {Gilliland}}]{Charbonneau02}
{Charbonneau}, D., {Brown}, T.~M., {Noyes}, R.~W., \& {Gilliland}, R.~L. 2002,
  \apj, 568, 377

\bibitem[{{Chen} \& {Kipping}(2017)}]{Chen17}
{Chen}, J. \& {Kipping}, D. 2017, \apj, 834, 17

\bibitem[{{Chilcote} {et~al.}(2017){Chilcote}, {Pueyo}, {De Rosa}, {Vargas},
  {Macintosh}, {Bailey}, {Barman}, {Bauman}, {Bruzzone}, {Bulger}, {Burrows},
  {Cardwell}, {Chen}, {Cotten}, {Dillon}, {Doyon}, {Draper}, {Duch{\^e}ne},
  {Dunn}, {Erikson}, {Fitzgerald}, {Follette}, {Gavel}, {Goodsell}, {Graham},
  {Greenbaum}, {Hartung}, {Hibon}, {Hung}, {Ingraham}, {Kalas}, {Konopacky},
  {Larkin}, {Maire}, {Marchis}, {Marley}, {Marois}, {Metchev},
  {Millar-Blanchaer}, {Morzinski}, {Nielsen}, {Norton}, {Oppenheimer},
  {Palmer}, {Patience}, {Perrin}, {Poyneer}, {Rajan}, {Rameau},
  {Rantakyr{\"o}}, {Sadakuni}, {Saddlemyer}, {Savransky}, {Schneider}, {Serio},
  {Sivaramakrishnan}, {Song}, {Soummer}, {Thomas}, {Wallace}, {Wang},
  {Ward-Duong}, {Wiktorowicz}, \& {Wolff}}]{Chilcote17}
{Chilcote}, J., {Pueyo}, L., {De Rosa}, R.~J., {et~al.} 2017, \aj, 153, 182

\bibitem[{{Fabrycky} {et~al.}(2014){Fabrycky}, {Lissauer}, {Ragozzine}, {Rowe},
  {Steffen}, {Agol}, {Barclay}, {Batalha}, {Borucki}, {Ciardi}, {Ford},
  {Gautier}, {Geary}, {Holman}, {Jenkins}, {Li}, {Morehead}, {Morris},
  {Shporer}, {Smith}, {Still}, \& {Van Cleve}}]{Fabrycky14}
{Fabrycky}, D.~C., {Lissauer}, J.~J., {Ragozzine}, D., {et~al.} 2014, \apj,
  790, 146

\bibitem[{{Fischer} {et~al.}(2002){Fischer}, {Marcy}, {Butler}, {Laughlin}, \&
  {Vogt}}]{Fischer02}
{Fischer}, D.~A., {Marcy}, G.~W., {Butler}, R.~P., {Laughlin}, G., \& {Vogt},
  S.~S. 2002, \apj, 564, 1028

\bibitem[{{Gillon} {et~al.}(2017){Gillon}, {Triaud}, {Demory}, {Jehin}, {Agol},
  {Deck}, {Lederer}, {de Wit}, {Burdanov}, {Ingalls}, {Bolmont}, {Leconte},
  {Raymond}, {Selsis}, {Turbet}, {Barkaoui}, {Burgasser}, {Burleigh}, {Carey},
  {Chaushev}, {Copperwheat}, {Delrez}, {Fernandes}, {Holdsworth}, {Kotze}, {Van
  Grootel}, {Almleaky}, {Benkhaldoun}, {Magain}, \& {Queloz}}]{Gillon17}
{Gillon}, M., {Triaud}, A.~H.~M.~J., {Demory}, B.-O., {et~al.} 2017, \nat, 542,
  456

\bibitem[{{Koen}(2003)}]{Koen03a}
{Koen}, C. 2003, \mnras, 341, 1385

\bibitem[{{Koen} {et~al.}(2003){Koen}, {Balona}, {Khadaroo}, {Lane},
  {Prinsloo}, {Smith}, \& {Laney}}]{Koen03b}
{Koen}, C., {Balona}, L.~A., {Khadaroo}, K., {et~al.} 2003, \mnras, 344, 1250

\bibitem[{{Kov{\'a}cs} {et~al.}(2002){Kov{\'a}cs}, {Zucker}, \&
  {Mazeh}}]{Kovacs02}
{Kov{\'a}cs}, G., {Zucker}, S., \& {Mazeh}, T. 2002, \aap, 391, 369

\bibitem[{{Kreidberg}(2015)}]{Kreidberg15}
{Kreidberg}, L. 2015, \pasp, 127, 1161

\bibitem[{{Kuschnig} {et~al.}(1997){Kuschnig}, {Weiss}, {Gruber}, {Bely}, \&
  {Jenkner}}]{kuschnig1997}
{Kuschnig}, R., {Weiss}, W.~W., {Gruber}, R., {Bely}, P.~Y., \& {Jenkner}, H.
  1997, \aap, 328, 544

\bibitem[{{Lagrange} {et~al.}(2010){Lagrange}, {Bonnefoy}, {Chauvin}, {Apai},
  {Ehrenreich}, {Boccaletti}, {Gratadour}, {Rouan}, {Mouillet}, {Lacour}, \&
  {Kasper}}]{Lagrange10}
{Lagrange}, A.-M., {Bonnefoy}, M., {Chauvin}, G., {et~al.} 2010, Science, 329,
  57

\bibitem[{{Lagrange} {et~al.}(2009){Lagrange}, {Gratadour}, {Chauvin}, {Fusco},
  {Ehrenreich}, {Mouillet}, {Rousset}, {Rouan}, {Allard}, {Gendron}, {Charton},
  {Mugnier}, {Rabou}, {Montri}, \& {Lacombe}}]{Lagrange09}
{Lagrange}, A.-M., {Gratadour}, D., {Chauvin}, G., {et~al.} 2009, \aap, 493,
  L21

\bibitem[{{Lagrange} {et~al.}(2013){Lagrange}, {Meunier}, {Chauvin}, {Sterzik},
  {Galland}, {Lo Curto}, {Rameau}, \& {Sosnowska}}]{Lagrange13}
{Lagrange}, A.-M., {Meunier}, N., {Chauvin}, G., {et~al.} 2013, \aap, 559, A83

\bibitem[{{Larwood} \& {Kalas}(2001)}]{Larwood01}
{Larwood}, J.~D. \& {Kalas}, P.~G. 2001, \mnras, 323, 402

\bibitem[{{Lenz} \& {Breger}(2005)}]{lenz2005}
{Lenz}, P. \& {Breger}, M. 2005, Communications in Asteroseismology, 146, 53

\bibitem[{{Mamajek} \& {Bell}(2014)}]{Mamajek14}
{Mamajek}, E.~E. \& {Bell}, C.~P.~M. 2014, \mnras, 445, 2169

\bibitem[{{Marcy} \& {Butler}(1995)}]{Marcy95}
{Marcy}, G.~W. \& {Butler}, R.~P. 1995, in Bulletin of the American
  Astronomical Society, Vol.~27, American Astronomical Society Meeting
  Abstracts, 1379

\bibitem[{{Marcy} {et~al.}(2001){Marcy}, {Butler}, {Fischer}, {Vogt},
  {Lissauer}, \& {Rivera}}]{Marcy01}
{Marcy}, G.~W., {Butler}, R.~P., {Fischer}, D., {et~al.} 2001, \apj, 556, 296

\bibitem[{{Mayor} \& {Queloz}(1995)}]{Mayor95}
{Mayor}, M. \& {Queloz}, D. 1995, \nat, 378, 355

\bibitem[{{Mouillet} {et~al.}(1997){Mouillet}, {Larwood}, {Papaloizou}, \&
  {Lagrange}}]{Mouillet08}
{Mouillet}, D., {Larwood}, J.~D., {Papaloizou}, J.~C.~B., \& {Lagrange}, A.~M.
  1997, \mnras, 292, 896

\bibitem[{{Pablo} {et~al.}(2016){Pablo}, {Whittaker}, {Popowicz}, {Mochnacki},
  {Kuschnig}, {Grant}, {Moffat}, {Rucinski}, {Matthews},
  {Schwarzenberg-Czerny}, {Handler}, {Weiss}, {Baade}, {Wade},
  {Zoc{\l}o{\'n}ska}, {Ramiaramanantsoa}, {Unterberger}, {Zwintz}, {Pigulski},
  {Rowe}, {Koudelka}, {Orlea{\'n}ski}, {Pamyatnykh}, {Neiner}, {Wawrzaszek},
  {Marciniszyn}, {Romano}, {Wo{\'z}niak}, {Zawistowski}, \& {Zee}}]{Pablo16}
{Pablo}, H., {Whittaker}, G.~N., {Popowicz}, A., {et~al.} 2016, \pasp, 128,
  125001

\bibitem[{{Pepper} {et~al.}(2007){Pepper}, {Pogge}, {DePoy}, {Marshall},
  {Stanek}, {Stutz}, {Poindexter}, {Siverd}, {O'Brien}, {Trueblood}, \&
  {Trueblood}}]{Pepper07}
{Pepper}, J., {Pogge}, R.~W., {DePoy}, D.~L., {et~al.} 2007, \pasp, 119, 923

\bibitem[{{Pigulski} {et~al.}(2016){Pigulski}, {Cugier}, {Popowicz},
  {Kuschnig}, {Moffat}, {Rucinski}, {Schwarzenberg-Czerny}, {Weiss}, {Handler},
  {Wade}, {Koudelka}, {Matthews}, {Mochnacki}, {Orlea{\'n}ski}, {Pablo},
  {Ramiaramanantsoa}, {Whittaker}, {Zoc{\l}o{\'n}ska}, \&
  {Zwintz}}]{pigulski2016}
{Pigulski}, A., {Cugier}, H., {Popowicz}, A., {et~al.} 2016, \aap, 588, A55

\bibitem[{{Popowicz}(2016)}]{Popowicz16}
{Popowicz}, A. 2016, in \procspie, Vol. 9904, Space Telescopes and
  Instrumentation 2016: Optical, Infrared, and Millimeter Wave, 99041R

\bibitem[{{Popowicz} {et~al.}(2017){Popowicz}, {Pigulski}, {Bernacki},
  {Kuschnig}, {Pablo}, {Ramiaramanantsoa}, {Zoc{\l}o{\'n}ska}, {Baade},
  {Handler}, {Moffat}, {Wade}, {Neiner}, {Rucinski}, {Weiss}, {Koudelka},
  {Orlea{\'n}ski}, {Schwarzenberg-Czerny}, \& {Zwintz}}]{popowicz2017}
{Popowicz}, A., {Pigulski}, A., {Bernacki}, K., {et~al.} 2017, \aap, 605, A26

\bibitem[{{Ricker} {et~al.}(2014){Ricker}, {Winn}, {Vanderspek}, {Latham},
  {Bakos}, {Bean}, {Berta-Thompson}, {Brown}, {Buchhave}, {Butler}, {Butler},
  {Chaplin}, {Charbonneau}, {Christensen-Dalsgaard}, {Clampin}, {Deming},
  {Doty}, {De Lee}, {Dressing}, {Dunham}, {Endl}, {Fressin}, {Ge}, {Henning},
  {Holman}, {Howard}, {Ida}, {Jenkins}, {Jernigan}, {Johnson}, {Kaltenegger},
  {Kawai}, {Kjeldsen}, {Laughlin}, {Levine}, {Lin}, {Lissauer}, {MacQueen},
  {Marcy}, {McCullough}, {Morton}, {Narita}, {Paegert}, {Palle}, {Pepe},
  {Pepper}, {Quirrenbach}, {Rinehart}, {Sasselov}, {Sato}, {Seager},
  {Sozzetti}, {Stassun}, {Sullivan}, {Szentgyorgyi}, {Torres}, {Udry}, \&
  {Villasenor}}]{Ricker14}
{Ricker}, G.~R., {Winn}, J.~N., {Vanderspek}, R., {et~al.} 2014, in \procspie,
  Vol. 9143, Space Telescopes and Instrumentation 2014: Optical, Infrared, and
  Millimeter Wave, 914320

\bibitem[{{Smith} \& {Terrile}(1984)}]{Smith84}
{Smith}, B.~A. \& {Terrile}, R.~J. 1984, Science, 226, 1421

\bibitem[{{Stuik} {et~al.}(2017){Stuik}, {Bailey}, {Dorval}, {Talens},
  {Laginja}, {Mellon}, {Lomberg}, {Crawford}, {Ireland}, {Mamajek}, \&
  {Kenworthy}}]{Stuik17}
{Stuik}, R., {Bailey}, J.~I., {Dorval}, P., {et~al.} 2017, \aap, 607, A45

\bibitem[{{Talens} {et~al.}(2017){Talens}, {Spronck}, {Lesage}, {Otten},
  {Stuik}, {Pollacco}, \& {Snellen}}]{Talens17}
{Talens}, G.~J.~J., {Spronck}, J.~F.~P., {Lesage}, A.-L., {et~al.} 2017, \aap,
  601, A11

\bibitem[{{van Leeuwen}(2007)}]{Leeuwen07}
{van Leeuwen}, F., ed. 2007, Astrophysics and Space Science Library, Vol. 350,
  {Hipparcos, the New Reduction of the Raw Data}

\bibitem[{{Wahhaj} {et~al.}(2003){Wahhaj}, {Koerner}, {Ressler}, {Werner},
  {Backman}, \& {Sargent}}]{Wahhaj03}
{Wahhaj}, Z., {Koerner}, D.~W., {Ressler}, M.~E., {et~al.} 2003, \apjl, 584,
  L27

\bibitem[{{Wang} {et~al.}(2016){Wang}, {Graham}, {Pueyo}, {Kalas},
  {Millar-Blanchaer}, {Ruffio}, {De Rosa}, {Ammons}, {Arriaga}, {Bailey},
  {Barman}, {Bulger}, {Burrows}, {Cardwell}, {Chen}, {Chilcote}, {Cotten},
  {Fitzgerald}, {Follette}, {Doyon}, {Duch{\^e}ne}, {Greenbaum}, {Hibon},
  {Hung}, {Ingraham}, {Konopacky}, {Larkin}, {Macintosh}, {Maire}, {Marchis},
  {Marley}, {Marois}, {Metchev}, {Nielsen}, {Oppenheimer}, {Palmer}, {Patel},
  {Patience}, {Perrin}, {Poyneer}, {Rajan}, {Rameau}, {Rantakyr{\"o}},
  {Savransky}, {Sivaramakrishnan}, {Song}, {Soummer}, {Thomas}, {Vasisht},
  {Vega}, {Wallace}, {Ward-Duong}, {Wiktorowicz}, \& {Wolff}}]{Wang16}
{Wang}, J.~J., {Graham}, J.~R., {Pueyo}, L., {et~al.} 2016, \aj, 152, 97

\bibitem[{{Weiss} {et~al.}(2014){Weiss}, {Rucinski}, {Moffat},
  {Schwarzenberg-Czerny}, {Koudelka}, {Grant}, {Zee}, {Kuschnig}, {Mochnacki},
  {Matthews}, {Orleanski}, {Pamyatnykh}, {Pigulski}, {Alves}, {Guedel},
  {Handler}, {Wade}, \& {Zwintz}}]{Weiss14}
{Weiss}, W.~W., {Rucinski}, S.~M., {Moffat}, A.~F.~J., {et~al.} 2014, \pasp,
  126, 573

\bibitem[{{Zwintz}(2017)}]{Zwintz17}
{Zwintz}, K. 2017, ArXiv e-prints

\end{thebibliography}
\end{document}